\documentclass[a4paper,11pt]{article}
\usepackage{graphicx}
\usepackage{xcolor}
\usepackage{amsmath,amsfonts,amssymb,amstext,hyperref,hypcap}
\usepackage{cancel}
\usepackage{empheq}
\usepackage{float}
\usepackage{cite}
\usepackage{subcaption}
\usepackage{authblk}
\usepackage{comment}
\usepackage{blindtext}
\usepackage{cases}
\usepackage{ulem}
\usepackage{epsfig}%
\usepackage[bottom]{footmisc}

\usepackage{epsfig}%
\usepackage[bottom]{footmisc}
\begin{document}
\date{}

\title{\centerline \bf Observational constraints on non-minimally coupled curvature-matter models of gravity from the analysis of Pantheon data}
\bigskip

\author[1]{Biswajit Jana\thanks{vijnanachaitanya2020@gmail.com}} 
\author[2]{Anirban Chatterjee \thanks{Corresponding Author: anirbanc@iitk.ac.in \& iitkanirbanc@gmail.com}}
\author[1]{Kumar Ravi \thanks{cimplyravi@gmail.com}}
\author[1]{Abhijit Bandyopadhyay\thanks{abhijit.phy@gm.rkmvu.ac.in}}
\normalsize
\affil[1]{Ramakrishna Mission Vivekananda Educational and Research Institute, 
Belur Math, Howrah 711202, India}
\affil[2]{Indian Institute of Technology Kanpur, Kanpur 208016, India}
\date{\today}
\maketitle

\begin{abstract}
We considered non-minimally coupled curvature-matter models of gravity in a FRW universe filled with perfect fluid and investigated its cosmological implications in the light of Pantheon compilation of 1048 Supernova Ia data points along with 54 data points from Observed Hubble Data. The non-minimal curvature-matter coupling has been introduced by adding a term $\int  \left[\lambda R^n \mathcal{L}_m\right] \sqrt{-g}d^4x$ to the usual action for Einstein gravity involving the Einstein Hilbert action and minimally coupled matter action.
We investigate the observational constraints on the non-minimal models by choosing two different kinds of parametrization of fluid-pressure profiles using a dimensionless parameter $k$. The interplay of the three parameters $\lambda,n,k$ plays a pivotal role in testing the consistency of non-minimally coupled fluid-curvature scenarios with the observed data.  We found there exist large domains in the ($\lambda,n,k$)-parameter space for which models with non-minimal curvature-matter mixing stand as viable cosmological models reproducing the observed features of late-time cosmic evolution. We also commented on the possibility of `gravitationally induced particle creation' in the context of SNe Ia data. 
\end{abstract}

\section{Introduction}
Observation of type Ia Supernovae (SNe Ia) events
and their luminosity distance and redshift measurements by  Riess \textit{et.al.} \cite{ref:Riess98} and Perlmutter \textit{et.al.}  
\cite{ref:Perlmutter} have independently confirmed
that present universe is in an accelerated state
of expansion and a  transition from a 
decelerated to the current accelerated phase happened during 
it's late time phase of evolution.
The reason for this late-time cosmic acceleration is attributed to a hypothetical unclustered form of energy, called 
`Dark Energy' (DE), which has become 
a general label for the late time cosmic acceleration.
On the other hand, observations of rotation curves of spiral galaxies \cite{Sofue:2000jx}, 
gravitational lensing \cite{Bartelmann:1999yn},
Bullet Cluster \cite{Clowe:2003tk}  and other colliding 
clusters provide evidence in favour of 
existence of non-luminous matter called `Dark Matter' (DM) in the universe. Such matter 
indirectly manifest their existence only through gravitational interactions.
Experiments like WMAP \cite{Hinshaw:2008kr} and Planck \cite{Ade:2013zuv} 
have revealed that   dark energy (DE) and dark matter (DM) together 
contribute $\sim 96\%$ of total energy density of the present-day universe, 
with  $\sim$69$\%$  and $\sim$27$\%$ as their respective shares.  
The rest of the 4$\%$ contribution comes from radiation and baryonic matter.    \\

To support  his theory of the static universe, Einstein
 initially considered a cosmological constant term $\Lambda g_{\mu\nu}$  in the geometric part of his field equations of general relativity, but later discarded it, in favour of Hubble's observation of expanding universe. However, with the discovery of late-time cosmic acceleration 
in the later part of last century,
the cosmological constant term again   came  under the spotlight 
because of its  potential to provide a straight-way solution to
Einstein's field equation resulting in an accelerated expansion.   
The corresponding phenomenological model that fits the observed features  
of cosmic acceleration is called
$\Lambda$-CDM model, where CDM refers to cold dark matter content of the present universe.
Unfortunately, the $\Lambda$-CDM model is plagued with the coincidence problem \cite{Zlatev:1998tr} problem and the fine-tuning problem \cite{Martin:2012bt}. These limitations
motivate investigation of other models to account for dark energy.
One class of such models
are field theoretic models of dark energy involving
modification of energy-momentum tensor in the Einstein's field
equations due to presence of a field as one content of the 
universe, other than matter and radiation. Such class of models
include both quintessence   \cite{Peccei:1987mm,Ford:1987de,Peebles:2002gy,Nishioka:1992sg, Ferreira:1997au,Ferreira:1997hj,Caldwell:1997ii,Carroll:1998zi,Copeland:1997et},
and $k$-essence   \cite{Fang:2014qga,ArmendarizPicon:1999rj,ArmendarizPicon:2000ah,ArmendarizPicon:2000dh,ArmendarizPicon:2005nz,Chiba:1999ka,ArkaniHamed:2003uy,Caldwell:1999ew} models of scalar fields. Another class of models involve  modification of the
geometric part of Einstein's equations, \textit{i.e.} the Einstein-Hilbert action
in order to address issue of late time cosmic acceleration. Such models include 
$f(R)$ gravity models, scalar-tensor theories, Gauss-Bonnet gravity, and braneworld models of dark energy \cite{fr1}. In most of these scenarios, the  universe is considered to
be homogeneous and isotropic at large scales, with
its metric described by the Friedmann-Robertson-Walker (FRW) metric
specified by a time-dependent scale factor $a(t)$ and a curvature constant $K$.
There also exist approaches,
where an inhomogeneous universe, described by a perturbed FRW metric,
has been considered as the spacetime background. \\

In this article, we examine a model of the universe 
with its content as  
an ideal hypothetical fluid, characterised by
its energy-density $(\rho)$ and pressure $(p)$, 
which is non-minimally coupled
to the spacetime curvature.
The non-minimal curvature-matter coupling has been introduced
by adding a term  $\int d^4x \sqrt{-g} \lambda R^n {\cal L}_m$ to the usual action for Einstein gravity involving the
Einstein Hilbert action and minimally coupled matter action.
$R$ is the Ricci Scalar and ${\cal L}_m$ (referred to as
matter Lagrangian)   
is the Lagrangian of the hypothetical fluid. 
The coupling constant ($\lambda$) and the power ($n$) 
of $R$ in the  non-minimal  coupling term, are
considered as the parameters of the model.
We explored the constraints on these parameters  
from the combined analysis of
 Pantheon compilation of 1048 SNe Ia data points \cite{Pan-STARRS1:2017jku} and 54 data
points  from Observed Hubble data \cite{Blake, Chuang, Font-Ribera, Delubac, Bautista}.\\

In the general framework of non-minimal models, the 
geometric part of Einstein's equation is altered  
by modifying the Einstein-Hilbert action to $\int d^4x \sqrt{-g} f(R)$,
replacing $R$ by a general function $f(R)$ 
\cite{Bertolami:2008ab,Schutz:1970,Brown:1993}. 
In the context of  modified $f(R)$ theories of
gravity, it has been   shown in \cite{Bertolami:2007} that the covariant derivative of energy-momentum tensor is non-vanishing  
($\nabla_{\mu}T^{\mu \nu} \neq 0$),
when a $f(R)$-${\cal L}_m$ (curvature-matter) coupling is present 
in the theory which may cause a departure from geodesic motion  manifesting
existence of a  new force. Consequences of such models in 
stellar equilibrium  have been investigated in 
\cite{Bertolami:2007vu,Sotiriou:2008dh}. An equivalence between
a suitable scalar theory and  the generic model with
non-minimal coupling between curvature and matter 
have been established in \cite{Bertolami:2008im, Azevedo1, Azevedo2, Azevedo3, Azevedo4, Sen, Tiwari:2023vcc}.
A  viability criterion to avoid instabilities in such models has
been derived in  \cite{Faraoni:2007sn}. 
The effect of curvature-matter couplings on the dynamics   of particles
and fields have been analysed in \cite{Sotiriou:2008it}. 
In \cite{Bertolami:2007vu,Sotiriou:2008dh}, 
it has   been argued that the choice
${\cal L}_m = p$, ($p$ being the pressure
of the matter fluid) is a `natural' one  for the matter of Lagrangian density,
as it reproduces the equations of hydrodynamics of perfect fluid
and in the context of curvature-matter coupling such choices result in
vanishing of the extra force. However,  
there may be some other choices resulting in vanishing the extra force (see  \cite{Brown:1993,Hawking:1973} for references).  In this work
we have used ${\cal L}_m = p$ and in the framework of our considerations 
the field
equations following from the total action involving
$S_{\rm non-minimal}$ establishes connection
between $a(t), \rho(t), p(t)$ through their time derivatives and various other
functions like the Hubble function $H(t)$ and the Ricci scalar $R(t)$.
We exploit these equations
for investigating
constraints on non-minimal models from the analysis of observed data.\\

Using a model-independent construction
of Hubble's function $H$ in terms of redshift $z$
without invoking any specific cosmological models,
we obtained the dependence of  Hubble's 
parameter $H$ on $z$ from the analysis of
the SNe Ia data (Pantheon + OHD). We used the FRW scale factor $a$
to be normalised to unity at 
present epoch ($z=0$).  Using the relation $1/a = 1+z$,
the temporal behaviour of cosmological quantities
expressed in terms of $z$ 
  may be equivalently expressed in terms of $a$ or 
  another dimensionless time parameter 
chosen as $\tau \equiv \ln a$, which we used in our analysis.
Exploiting the profile of the Hubble's parameter for the redshift domain accessible in the SNe Ia observations, obtained from the analysis of cosmological data set, we obtain the time profile of FRW scale factor $a$ and it's higher-order derivatives $\dot{a}$, $\ddot{a}$.  This also enables us to express temporal behaviour of the Ricci scalar $R$ and it's time derivatives which are involved in our analysis.\\

To investigate the observational constraints on models
with non-minimal  curvature-matter couplings (specified
in terms of parameters $\lambda$ and $n$ in this paper),
we considered two different types of fluid pressure models
with temporal behaviour of pressure $p$ modelled as $p \sim e^{ak}$ (exponential model) 
and $p \sim a^k$ (power law model) where $k$ is a dimensionless parameter.
Consequently  the three parameters ($\lambda,n,k$) 
appear in the evolution equations and control
the model-based computations of $\rho$ and $p$.
Using the observational inputs from SNe Ia data, 
we obtained  the regions in the model parameter space ($\lambda,n,k$),
for which the computed energy density ($\rho$)  of the fluid
remains positive for all time during the late time phase of evolution.
This is an essential requirement for viability
of cosmological models.  We have also seen that there
exist a small range of parameter values around $(\lambda = -0.1, n = 0.2, k = 1)$
 for which the computed temporal profiles of $\rho$ and $p$ 
 mimic the corresponding profiles obtained from the analysis
of the data using $\Lambda$-CDM model in the context of usual minimal coupling scenario.\\

The non-vanishing covariant derivative of the energy-momentum tensor in the context
of non-minimally coupled curvature-matter scenarios implies an exchange of energy
between curvature and matter sectors.
We computed 
the rate of the energy exchange for different benchmark values
of the model parameters ($\lambda,n,k$)  
and found a monotonous decrease in the absolute value of the 
energy exchange rate as time approaches towards the present epoch.
This  implies that the 
rate of exchange of energy between the two sectors
is more significant during relatively early phases of late-time cosmic
evolution.
The cosmological implications of
curvature-matter coupling scenarios from the viewpoint of thermodynamics have been investigated in
\cite{Harko:2015pma,Harko:2014pqa,Moraes:2016mlp}. There it has been shown that
curvature-matter coupling may be responsible for generation of a large amount of
comoving  entropy during late-time evolutionary
phase of the universe leading  to a possible interpretation of
the exchange of energy between curvature and matter sector 
in terms of gravitationally induced particle creation in FRW universe.
We have shown from the analysis of SNe Ia data that for some values of the
model parameters ($\lambda,n,k$) possibility of such an interpretation is always allowed. \\

Rest of the paper is arranged in the following manner.
In  Sec.\ \ref{Sec:II} we   discussed the framework of
non-minimally coupled $f(R)$ model  in the context of a flat FRW universe 
and obtained the corresponding generic modified evolution
 equations of the universe in presence of non-minimal curvature-matter coupling. 
We choose two different types of fluid-pressure models,
and obtained the equations corresponding to each of the models.
 In Sec.\ \ref{Sec:III} we discussed the  methodology of analysis of Pantheon + OHD 
 data sets for obtaining temporal 
 behaviour of different relevant cosmological
quantities during the late-time phase of cosmic evolution.
In Sec. \ref{Sec:IV} we presented  the constraints on model parameters $\lambda$, 
 $n$ and $k$ obtained
by using the results extracted from the analysis of the observed data 
and discussed the results.
We summarize the conclusions of the paper in Sec.\ \ref{Sec:V}.

\section{Theoretical framework of curvature-matter coupling scenario}
\label{Sec:II}
In the framework of  modified  theories of gravity with non-minimal
curvature-matter coupling, the action may be written in the form 
\cite{Bertolami:2008ab} 
\begin{eqnarray}
\label{model} S &=& \int \left[{1 \over 2}f_1(R)+\left[1+\lambda
f_2(R)\right]{\cal L}_{m}\right] \sqrt{-g}\;d^{4}x\,,
\label{eq:b1}
\end{eqnarray}
where $f_1(R)$ and $f_2(R)$, in general,  are two arbitrary functions 
of the Ricci scalar $R$ and ${\cal L}_{m}$ is the  Lagrangian
density for matter. $g$ is the determinant of the spacetime metric tensor $g_{\mu\nu}$
and $\lambda$ is a constant representing
the strength of the non-minimal coupling between curvature term $f_2(R)$
and matter lagrangian ${\cal L}_m$.
Following the approach of metric formalism in the context of the modified
theories of gravity, the
variation of  the action with respect to 
the field $g_{\mu \nu }$ yields the modified field equations as 
\begin{eqnarray}
F_1R_{\mu \nu }-{1\over 2}f_1g_{\mu \nu }-\nabla_\mu \nabla_\nu
F_1+g_{\mu\nu}\square F_1 &=& (1+\lambda f_2)T_{\mu
\nu} -2\lambda F_2{\cal L}_m R_{\mu\nu}    \nonumber \\
&& + 2\lambda(\nabla_\mu
\nabla_\nu-g_{\mu\nu}\square){\cal L}_m F_2 \,,
\label{eq:b2}
\end{eqnarray}
where we put $8\pi G = 1$ ($G$ = Gravitational Constant), $F_i = df_i/dR$ ($i=1,2$)  and 
$T_{\mu \nu}$ is the matter energy-momentum tensor  given by 
\begin{eqnarray}
T_{\mu \nu}=-{2 \over \sqrt{-g}}{\delta(\sqrt{-g}{\cal
L}_m)\over \delta(g^{\mu\nu})} \,.
\label{eq:b3}
\end{eqnarray}

In this article, we  investigated the scenario of non-minimal coupling
between curvature and matter with geometric part of field equation
driven by pure Einstein gravity \textit{i.e.} with $f_1(R)=R$
and considered the matter part to be non-minimally coupled (with coupling
strength $\lambda$) to curvature through the function $f_2(R) = R^n$ 
(where $n$ is some constant).
In this model, the matter part of the universe is
described by a perfect fluid 
characterised by energy density $\rho$ and pressure $p$.  
Following the comprehensive discussion in  \cite{Schutz:1970, Brown:1993},
we assume $\mathcal{L}_m = p$, which is a `natural choice' for
Lagrangian density for perfect fluids, which correctly
reproduces the hydrodynamical equations for a perfect fluid.
This particular choice has interesting consequences in the 
analysis of curvature-matter coupling scenarios implying
vanishing of extra force owing
to departure from motion along the geodesic which may arise due to 
non-vanishing covariant derivative of $T_{\mu\nu}$ in the context of
curvature-matter coupling \cite{Bertolami:2007}. There are, however, other
choices of $\mathcal{L}_m$ investigated in \cite{Bertolami:2008ab, Brown:1993}. \\

We conducted an   exploration of the  intricacies associated
with non-minimal coupling between the curvature  and matter  by employing 
a power law form of the function  $f_2(R) =  R^n$, where $n$ is a constant exponent. Apart from the strength of the coupling
$\lambda$,  such consideration introduces a 
single  additional parameter $n$ into the analytical framework. 
 These two parameters $(\lambda, n)$, along with
a third parameter $k$, used to express temporal variation of 
the fluid pressure, constitute a   3-dimensional parameter space
$(\lambda, n, k)$ facilitating the exploration of 
 non-minimally coupled theories with fluids 
 having a wide range of pressure variation modes, with an optimum number of parameters. 
 It's important to recognize that the power law representation $f_2(R) = R^n$ for any real  $n$ may be   represented through the expansion of $f_2(R)$ around a non-zero reference point $R=R_0$ in the form of  infinite series: $f_2(R) =\sum_m c_m (R-R_0)^m$, where $m$ is an integer,   encompassing negative values as well, when $n$ is negative.
Nevertheless, the scenario of non-minimal couplings may theoretically be explored using particular $f_2(R)$ expressions  consisting of a finite number of terms incorporating diverse integral powers (both positive and negative) of $R$.  
Consideration of such forms  would then introduce
 more parameters (the coefficients of the different
integral powers of $R$) into the analysis scheme compared to the power-law version  of $f_2(R)$. 
Under such considerations, obtaining observational constraints on non-minimally coupled scenarios necessitate  exploring parameter spaces of significantly higher dimensions.
This also  
requires  assessment of distinct parameter combinations  for each specific expression of $f_2(R)$ to encompass the wide spectrum of $f_2(R)$ patterns. In contrast, opting for the power law form $f_2(R) = R^n$ (as adopted for this study), covers 
an alternative array of varied $f(R)$ profiles (demonstrating either monotonic increase or decrease)  achieved solely by modifying the exponent  $n$. This choice renders the investigation more manageable in terms of the triad $(\lambda, n, k)$.  
We have also come to recognize that the adoption of the power law form $f_2(R) = R^n$
 has proven advantageous in terms of  computational time and efficiency,
in pinpointing the precise value of $n$ for which the evaluated temporal profile of energy density and fluid pressure mimic  the corresponding patterns seen in the $\Lambda$-CDM model. To summarise our considerations for the present analysis, we have chosen 
 $f_1(R) =R,  f_2(R) = R^n, F_2(R) = nR^{n-1},  {\cal L}_m = p$. For such choices Eq. \eqref{eq:b2} can be expressed as
\begin{eqnarray}
R_{\mu \nu }- {1\over 2}Rg_{\mu \nu }
&=& 
(1+\lambda R^n)T_{\mu\nu} -2\lambda (nR^{n-1}) pR_{\mu\nu}
 +2\lambda(\nabla_\mu
\nabla_\nu-g_{\mu\nu}\square)p(nR^{n-1})\,.
\label{eq:b4}
\end{eqnarray}
The energy density $\rho$ and pressure $p$ of the fluid are respectively obtained from `$00$' and `$ii$' components
of the energy-momentum tensor $T_{\mu\nu}$. The `$00$' component of Eq.\ (\ref{eq:b4}) is 
\begin{eqnarray}
R_{00} - {1\over 2}Rg_{00}
&=& 
(1+\lambda R^n)T_{00} -2\lambda p (nR^{n-1})R_{00} 
- 6\lambda H (\dot{p}nR^{n-1} + pn(n-1)R^{n-2}\dot{R})\,.
\label{eq:newrho1}
\end{eqnarray}
For a FRW spacetime background the above equation takes a form which after some rearrangements may be written as 
\begin{eqnarray}
\rho &=& \frac{1}{1+\lambda R^n}\left[3H^2 +6n\lambda\{(n-1)HR^{n-2}\dot{R} -  (H^2 + \dot{H})R^{n-1}\}p +6n\lambda HR^{n-1} \dot{p} \right]\,.
\label{eq:rhor^n}
\end{eqnarray}
On the other hand, the  `$ii$'-component of Eq.\ (\ref{eq:b4}) is
\begin{eqnarray}
R_{11}- {1\over 2}Rg_{11}
&=& 
(1+\lambda R^n)T_{11} -2\lambda pnR^{n-1}R_{11} + 2\lambda(\nabla_1
\nabla_1 -g_{11}\square)(pnR^{n-1}) \,,
\label{eq:newp} 
\end{eqnarray}
which in the FRW spacetime background takes the form
\begin{eqnarray}
- (2\dot{H} + 3H^2) &=& \Big{[}(1+\lambda R^n -2\lambda nR^{n-1} (\dot{H} + 3H^2) +4n(n-1)\lambda H R^{n-2}\dot{R}\nonumber \\
&& +2\lambda \{n(n-1)(n-2)R^{n-3}\dot{R}^2 +n(n-1)R^{n-2}\ddot{R}\}\Big{]}p \nonumber \\
&& +4n\lambda \Big{\{} HR^{n-1} + (n-1)R^{n-2}\dot{R}\Big{\}}\dot{p} +2n\lambda R^{n-1}\ddot{p}  \,.
\label{eq:pr^n} 
\end{eqnarray}
In the framework of curvature-matter coupling in FRW spacetime background 
Eqs.   \eqref{eq:rhor^n} and \eqref{eq:pr^n} are the master equations. 
Note that Eqs.  \eqref{eq:rhor^n} and \eqref{eq:pr^n} involve time derivatives of the pressure of
the fluid which is a consequence of the non-minimal curvature-matter coupling. 
Setting the non-minimal coupling constant $\lambda = 0$ in the master equation, we retrieve the
usual Friedmann equations which do not contain any derivative term of pressure. \\

To investigate the observational constraints on such  
curvature-matter coupling models, we 
considered two types of fluid pressure models with 
specific temporal profile of the fluid pressure.
In  FRW background, we choose to express the modelled temporal profiles
of the fluid pressure in terms of the FRW scale factor. 
We discuss below the two models referred  to as exponential  and  power-law models as per the nature of 
dependences of the pressure on the scale factor in the corresponding models. \\

\textbf{Exponential model :} In this model we take the fluid pressure $p$ as $p=p_0\exp(ak)$, where $k$ is a dimensionless
parameter and  $p_0$ is a constant having the dimension of pressure. $p_0e^k$ gives the pressure of 
the fluid at present epoch $a=1$, in this model. We then have $\dot{p} = kp\dot{a}$ and $\ddot{p} = k^2p\dot{a}^2 + kp\ddot{a}$. 
Using $\dot{p} = kp\dot{a}$   in Eq. (\ref{eq:rhor^n}), 
 we may express the energy density for a given value of $\lambda,n,k$ and pressure $p$
corresponding to this model as
\begin{eqnarray}
\rho(t;\lambda,n,k) = \frac{1}{1+\lambda R^n}\left[3H^2 +6n\lambda\Big{\{}(n-1)HR^{n-2}\dot{R} -  (H^2 + \dot{H})R^{n-1} +kHR^{n-1}\dot{a}\Big{\}}p \right]\,.
\label{eq:rhoexp}
\end{eqnarray}
Note that the above equation also implies dependence of the $\rho(t;\lambda,n,k)$
on the parameter $p_0$ due to occurrence of the pressure term $p=p_0\exp(ak)$ in the
right hand side of Eq.\ (\ref{eq:rhoexp}). But note that, we may use 
  Eq. \eqref{eq:pr^n} with  $\dot{p} = kp\dot{a}$ and $\ddot{p} = k^2p\dot{a}^2 + kp\ddot{a}$ to express the pressure $p = p_0\exp(ak)$ as 
\begin{eqnarray}
p(t;\lambda,n,k) &=&- (2\dot{H} + 3H^2)\, \Big{[}(1+\lambda R^n -2\lambda nR^{n-1} (\dot{H} + 3H^2) +4n(n-1)\lambda H R^{n-2}\dot{R}\nonumber \\
&& +2\lambda \Big{\{}n(n-1)(n-2)R^{n-3}\dot{R}^2 +n(n-1)R^{n-2}\ddot{R}\Big{\}} \nonumber \\
&& +4n\lambda k \Big{\{} HR^{n-1} + (n-1)R^{n-2}\dot{R}\Big{\}}\dot{a} +2n\lambda k R^{n-1}(k\dot{a}^2 + \ddot{a})\Big{]}^{-1} \,.
\label{eq:pexp}
\end{eqnarray} 
We may now use this expression of Eq.(\ref{eq:pexp}) for $p$ in Eq.\ (\ref{eq:rhoexp})
to get rid of its explicit $p_0$ dependence. So that $\rho(t;\lambda,n,k)$ in Eq.\ (\ref{eq:rhoexp})
is   computable at any $t$ for  any given choice  of parameter set
$(\lambda,n,k)$.\\

\textbf{Power-law model :}
In this model we take   $p=p_0 a^k$, where $k$ is again a dimensionless constant and 
$p_0$ is a constant, which in this model gives the value of pressure at present epoch ($a=1$).
Here we thus have $\dot{p}=kpH$ and $\ddot{p}=p(k^2H^2+k\dot{H})$.
Using  $\dot{p}=kpH$  in   
Eq. (\ref{eq:rhor^n}) 
we may similarly express the energy density for a given set of values of $\lambda,n,k$ 
and pressure $p$ as
\begin{eqnarray}
\rho(t;\lambda,n,k) =\frac{1}{1+\lambda R^n}\left[3H^2 +6n\lambda\Big{\{}(n-1)HR^{n-2}\dot{R} -  (H^2 + \dot{H})R^{n-1} +kH^2R^{n-1}\Big{\}}p \right] \,. \label{eq:b6}
\end{eqnarray}
Here also, to get rid of the parameter $p_0$ which appears in the pressure term $p=p_0a^k$ in 
the right hand side, we exploit the
Eq. \eqref{eq:pr^n} with  $\dot{p} = kp\dot{a}$ and 
$\ddot{p} = k^2p\dot{a}^2 + kp\ddot{a}$ to express the instantaneous values of 
pressure $p = p_0a^k$ in terms of parameters  ($\lambda, n, k$)
as 
\begin{eqnarray}
p(t;\lambda,n,k) &=& - (2\dot{H} + 3H^2)\,\Big{[}(1+\lambda R^n -2\lambda nR^{n-1} (\dot{H} + 3H^2) +4n(n-1)\lambda H R^{n-2}\dot{R}\nonumber \\
&& +2\lambda \Big{\{}n(n-1)(n-2)R^{n-3}\dot{R}^2 +n(n-1)R^{n-2}\ddot{R}\Big{\}} \nonumber \\
&& +4n\lambda \Big{\{} HR^{n-1} + (n-1)R^{n-2}\dot{R}\Big{\}}kH +2n\lambda R^{n-1}(k^2H^2+k\dot{H})\Big{]}^{-1} 
\label{eq:b5}
\end{eqnarray}
and use this expression for pressure $p$ in Eq.\ (\ref{eq:b6})
to evaluate instantaneous value of energy density
for any given set of parameters $(\lambda,n,k)$.\\

Apart from the model parameters $(\lambda,n,k)$, the expressions
for $\rho$ and $p$ (in both exponential and power law models) involve
other cosmological parameters - the FRW scale factor $a$, its time derivatives, the Hubble parameter $H$ and its time 
derivatives, the Ricci scalar $R (= 12H^2 + 6\dot{H}$) and its time derivatives.
We have obtained the temporal behaviour of all these cosmological quantities 
during the late time phase of the cosmic evolution from a combined analysis of recently
released Pantheon (SNe Ia) data. The accessible range of red-shift
in the date sets corresponding to Pantheon data set is $0 \leq z \leq 2.3$
corresponds to a cosmic time domain $0.23 \leq t \leq 1$, where $t$ is normalised to
unity at the present epoch. Using the time dependences of the said cosmological
quantities as extracted from the observed data, we   obtained the
temporal behaviour of the energy density and pressure of the fluid
for both exponential and power law models using Eqs.\ (\ref{eq:rhoexp}), (\ref{eq:pexp}),  (\ref{eq:b6}) and  (\ref{eq:b5}). 
We obtained range of values of 
parameters $(\lambda,n,k)$ for which the evaluated values of energy density function $\rho(t;\lambda,n,k)$
remains positive at all time. The reach of the recent SNe Ia data accommodates within itself the scenario of matter-curvature non-minimal coupling for the constrained values of the model parameters. \\

\section{Obtaining constraints on the cosmological quantities from combined analysis of Pantheon and Observed Hubble data}
\label{Sec:III} 
The cosmological data from Supernovae Ia observation provides
information about late time cosmic evolution of the universe.
In this section we briefly outline the technical details of the analysis of the 
observational
data which involves Pantheon  (SNe Ia) data involving 1048
data points and Observed Hubble Data (OHD) involving 54 data points  
\cite{Blake, Chuang, Font-Ribera, Delubac, Bautista} and discussed
how we finally extracted the temporal behaviour of
the FRW scale factor $a$ and its derivatives which are essential
ingredients for constraining the non-minimally coupled matter-curvature
models as discussed in Sec.\ \ref{Sec:II}.\\

The SNe Ia data provides an estimation of luminosity distances ($d_L$)
at the redshift values  ($z$) corresponding to different  SNe Ia events.
The `Pantheon Sample' \cite{Pan-STARRS1:2017jku} 
is a compilation of the subset of 279 PS1 SNe Ia data points
(over a redshift range $0.03 < z < 0.68$) along with useful compilations of SNe Ia data
 from SDSS \cite{Frieman:2007mr}, SNLS \cite{SNLS:2005qlf}, various low-z \cite{cFa}, and HST samples  \cite{Riess:2006fw} 
  to form the largest combined sample of SNe Ia consisting of a total of 1048 SNe Ia  over the redshift range
$0.01 < z < 2.3$. We have used the above mentioned data samples to construct
the Hubble function $H(z)$  without invoking any specific cosmological models.
For such a model-independent construction, we use   Pad\'e approximant
of order (2,1)  \cite{G.A. Baker} for luminosity distance function,
which is a closed-form parametrization of the  luminosity distance in terms
of the redshift expressed as
\begin{eqnarray}
d_L (z,\alpha, \beta) = \frac{c}{H_0}\left( \frac{z(1+\alpha z)}{1+\beta z}\right) \,,
\label{eq:dL}
\end{eqnarray}
where $c$ is the speed of light and $H_0$ is the value of the Hubble parameter at the 
present epoch which may be defined through a dimensionless quantity $h$
by $H_0 = 100 h$ km s$^{-1}$ Mpc$^{-1}$. 
$\alpha$ and $\beta$ are the parameters to be determined from 
cosmological data.
The  Pantheon sample of 1048 data points provide
the redshift, apparent magnitude ($m$) at maximum brightness of the SNe Ia events, 
and also the covariance and correlations among the  different data points.
The observed data at any redshift $z_i$
 is expressed through the distance modulus $\mu_{\rm obs}(z_i)$
 which is given in terms of the absolute magnitude $M$
 and observed apparent magnitudes $m_{\rm obs}(z_i)$ by
 the relation
\begin{eqnarray}
\mu_{\rm obs}(z_i) &=& m_{\rm obs}(z_i) - M\,.
\label{eq:distmod1}
\end{eqnarray}
The form of the  
distance modulus $\mu_{\rm th}(z,\alpha,\beta)$ corresponding to the assumed parametric form of $d_L(z,\alpha,\beta)$ given in 
Eq.\ (\ref{eq:dL}) may be obtained by using the expression
\begin{eqnarray}
\mu_{\rm th}(z,\alpha,\beta)
&=&
5\log_{10} \left[\frac{H_0 d_L(z,\alpha,\beta)}{c}\right] + \mu_0
= 5\log_{10} \left[\frac{z(1+\alpha z)}{1+\beta z}\right] + \mu_0\,,
\label{eq:distmod2}
\end{eqnarray}
where $\mu_0 = 42.38 - 5\log_{10}h$. Computing   $\mu_{\rm th}(z_i,\alpha,\beta) - \mu_{\rm obs}(z_i)$ for the data points $i$, we may find the
$\chi^2$-function for the SN-data date defined by the relation
\begin{eqnarray}
\chi^2_0 
&=&
\sum_{i,j=1}^N 
\Big{[}\mu_{\rm th}(z_i,\alpha,\beta) - \mu_{\rm obs}(z_i)\Big{]}
C^{-1}_{ij} \Big{[}\mu_{\rm th}(z_j,\alpha,\beta) - \mu_{\rm obs}(z_j)\Big{]}\,,
\label{eq:chisqsn1}
\end{eqnarray}
where  $N = 1048$ is the total number of data points considered and $C$ is the covariance matrix of the data as released in \cite{Pan-STARRS1:2017jku}, which includes both statistical and
systematic uncertainties. Using Eqs.\ (\ref{eq:distmod1}) and (\ref{eq:distmod2}), the quantities $\mu_{\rm th}  - \mu_{\rm obs}$ appearing in Eq.\ (\ref{eq:chisqsn1}) may be expressed as
\begin{eqnarray}
\Big{[}\mu_{\rm th}(z_i,\alpha,\beta) - \mu_{\rm obs}(z_i)\Big{]}
&=&
5\log_{10} \left[\frac{z_i(1+\alpha z_i)}{1+\beta z_i}\right] - m_{\rm obs}(z_i) + M^\prime\,,
\label{eq:resd1}
\end{eqnarray}
where $M^\prime \equiv \mu_0 + M = 42.38 - 5\log_{10}h + M $. 
Since the Pantheon   data set comprises of apparent 
magnitudes only so this data-set alone cannot constrain $H_0$.
The parameter $H_0 (= 100 h$ km s$^{-1}$ Mpc$^{-1})$  enters in    
the $\chi^2$ function through the parameter $M^\prime$. 
Marginalising over the parameter $M^\prime$, called the  nuisance parameter, we define 
an appropriate $\chi^2$ function for this data set \cite{Abhi2012} as
\begin{eqnarray}
\chi^2_{\rm SN} &=& -2\ln \int_{-\infty}^\infty \exp\left[-\frac{1}{2}
\chi^2_0(\alpha,\beta,M^\prime)\right] dM^\prime \,,\label{eq:chisqsn2}
\end{eqnarray}
which may be put in the following form
\begin{eqnarray}
\chi^2_{\rm SN} &=& P  - \frac{Q^2}{R} + \ln (R/2\pi)\,,
\label{eq:chisqsn3}
\end{eqnarray}
where $P = (m_{\rm obs} - m_{\rm th})^TC^{-1}(m_{\rm obs} - m_{\rm th})$,
$Q = (m_{\rm obs} - m_{\rm th})^T C^{-1}\mathbf{1}$ and 
$R = \mathbf{1}^T C^{-1} \mathbf{1}$.
Here $(m_{\rm obs} - m_{\rm th})$
represents a column matrix of the residuals with $i^{\rm th}$ entry as $\Big{[}m_{\rm th}(z_i,\alpha,\beta) - m_{\rm obs}(z_i)\Big{]}$ and 
$(m_{\rm obs} - m_{\rm th})^T$ denotes the corresponding transposed matrix.
$m_{\rm obs}(z_i)$ is the observed value of apparent magnitude at redshift $z_i$
and corresponding theoretical value $m_{\rm th}(z)$ is given by the formula
$m_{\rm th}(z) = 5\log_{10} \left[\frac{z(1+\alpha z)}{1+\beta z}\right] + 25$.
The symbol $\mathbf{1}$ represents a column array of ones of same length as
$m_{\rm obs}$.\\

In our analysis we have also used 31 redshift \textit{vs} Hubble parameter
data points from chronometer 
observations \cite{Wei} and 23 data points from the line of sight 
Baryonic Acoustic Oscillations(BAO) data \cite{Blake, Chuang, 
Font-Ribera, Delubac, Bautista}. For a cosmological  model-independent construction of the Hubble parameter, we use the expression
\begin{eqnarray}
H(z,\alpha,\beta) = \left[\frac{1}{c}\frac{d}{dz}\left\{\frac{d_{L}(z,\alpha,\beta)}{(1+z)}\right\}\right]^{-1}\,,
\label{eq:hubble}
\end{eqnarray}
where $d_{L}(z,\alpha,\beta)$ is as given in Eq.\ (\ref{eq:dL}).
The residual for the Observed Hubble data involving total 54 data points
fitting is given by
\begin{eqnarray}
\chi^2_{\rm OHD} 
&=& \sum_{i=1}^{54} \left[ \frac{H(z_i,\alpha,\beta) - H_{\rm obs}(z_i)}{\sigma_i} \right]^2\,, 
\end{eqnarray}
where $H_{\rm obs}(z_i)$ and $\sigma_{i}$ give the observed
value of Hubble parameter and its corresponding uncertainty
 respectively  at redshift $z_i$
corresponding to the $i^{\rm th}$ data point of OHD data set.\\

\begin{figure}[H]	
\centering
\includegraphics[scale=0.55]{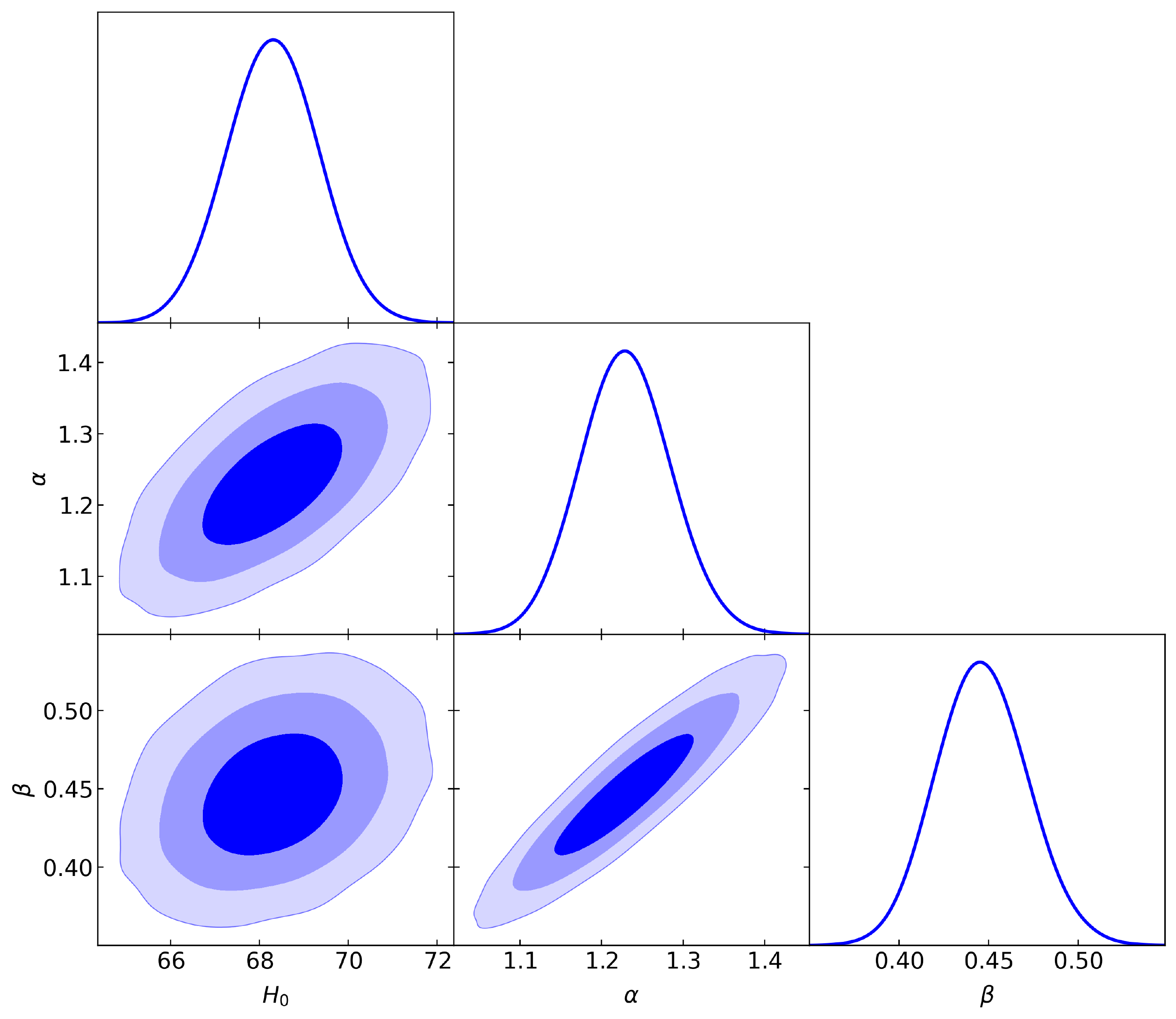} 
\caption{The posterior distribution plot of $H_0$, $\alpha$, and $\beta$}
\label{fig:Contour}
\end{figure}

To get estimates of the parameters $H_{0},\,\alpha,\,\beta$ along with their uncertainties, we perform  a
Markov chain Monte Carlo (MCMC) Bayesian parameter estimation
using a uniform prior to all the parameters.
We used the python packages, available in
the public domain, {\it emcee} \cite{Foreman} 
and   {\it GetDist} \cite{Lewis} for generating
 generate MCMC samples and for plotting posterior distributions of $H_0$, $\alpha$, and $\beta$  respectively. After performing  
various tests for the independence, convergence of the MCMC samples and thinning of the samples accordingly, we get following estimates from
the combined analysis of Pantheon data and OHD. In Fig.\ \ref{fig:Contour}
we presented the posterior distribution plot of $H_0$, $\alpha$, and $\beta$
which depicts the corresponding  
1$\sigma$ and 2$\sigma$ uncertainties for these parameters.
\begin{eqnarray}
\alpha = 1.23^{+0.06}_{-0.05},\quad \beta = 0.45^{+0.03}_{-0.03} , \quad {\rm and} \quad  H_0 = 68.31^{+1.03}_{-1.03} \mbox{ km s$^{-1}$ Mpc$^{-1}$}\,.
\end{eqnarray}

\begin{figure}[H]
\centering
\begin{subfigure}[b]{0.49\textwidth}
\includegraphics[width=\textwidth]{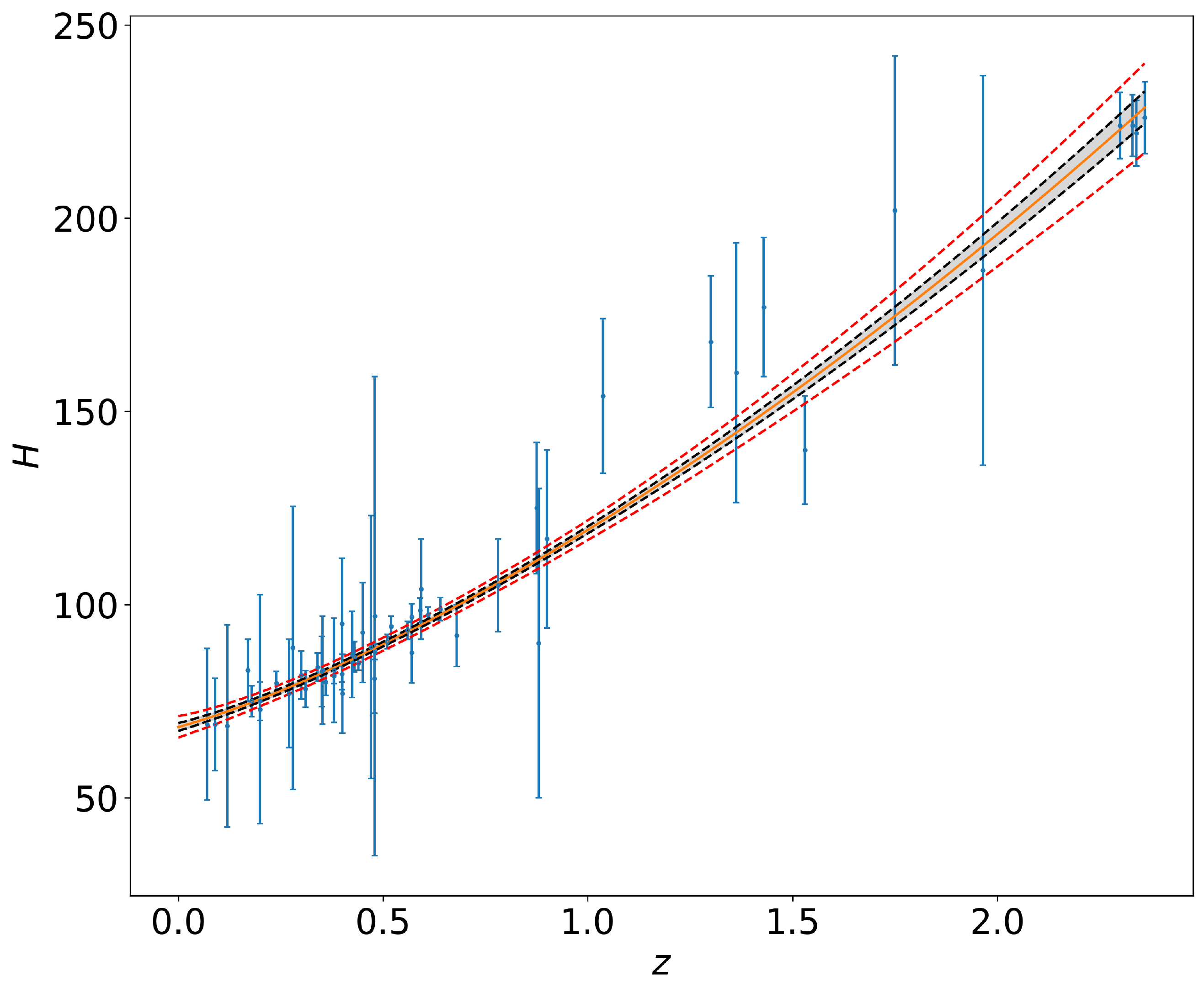}
\end{subfigure}
\begin{subfigure}[b]{0.49\textwidth}
\includegraphics[width=\textwidth]{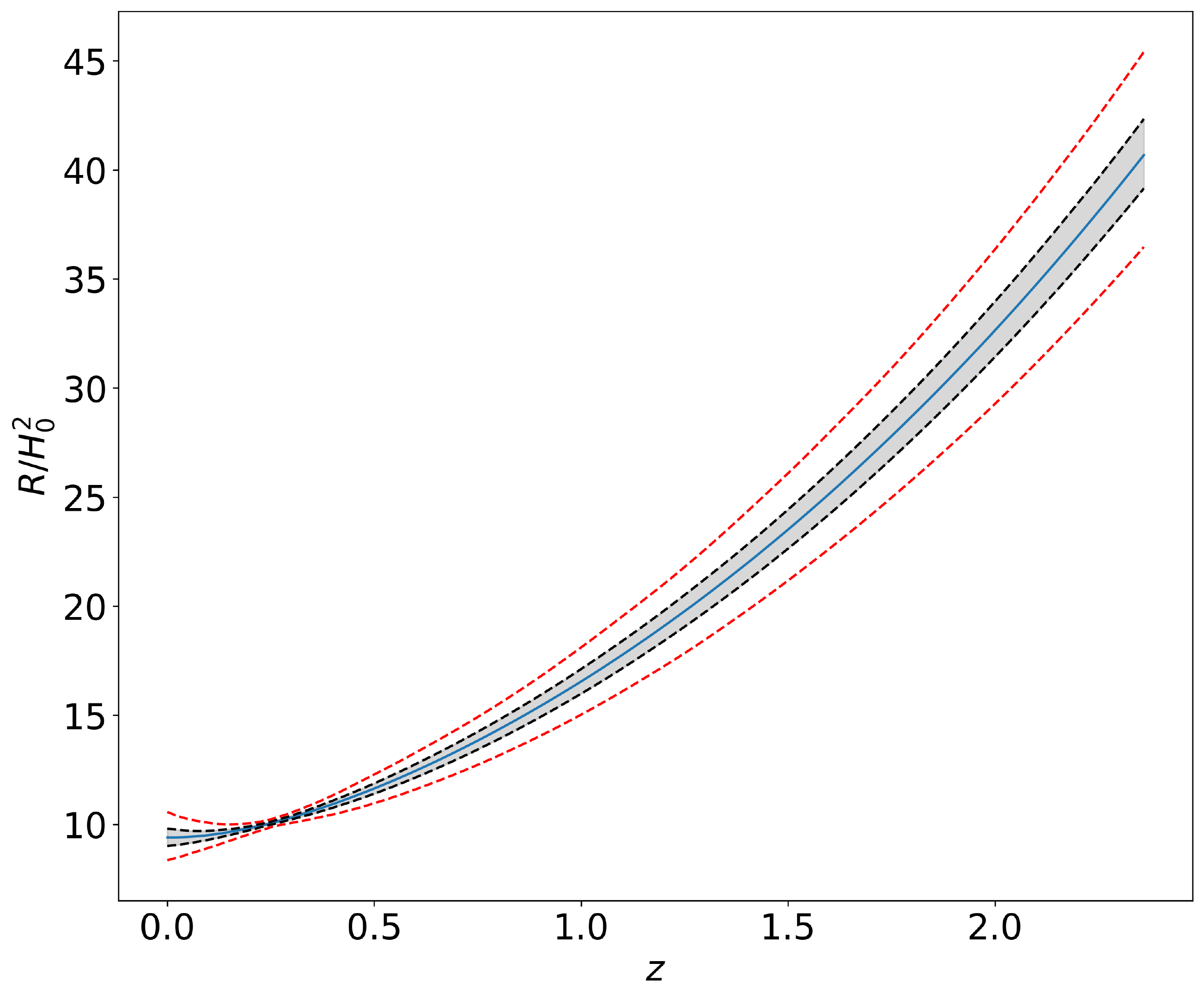}
\end{subfigure}
\caption{Left Panel: Best-fit curve with   1$\sigma$ and 3$\sigma$
uncertainties obtained for $H(z)$ vs $z$ from
analysis of Pantheon + OHD data. Observed values of $H$ (with error bars) corresponding
to the 54 data points  
from OHD are also shown. Right Panel:
Best-fit curve with 1$\sigma$ and 3$\sigma$
uncertainties obtained for $R(z)/H_0^2$ vs $z$ 
from the analysis} 
\label{fig:2}
\end{figure}

The obtained best-fit values of the parameters $\alpha$, $\beta$ and $H_0$
along with their respective uncertainties are then exploited to compute the 
redshift dependences (temporal profiles) of the 
Hubble parameter $H$ and the normalised Ricci Scalar  $R/H_0^2$ making
use of Eq.\ (\ref{eq:hubble}) and $R = 12H^2 + 6\dot{H}^2$.
We have shown the plots of $H$ and $R/H_0^2$ vs $z$ respectively in the left and right panels of  Fig.\ \ref{fig:2}.
The solid lines in  Fig.\ \ref{fig:2} correspond to the best-fit curve of each figure.
The $1\sigma$ and $3\sigma$ uncertainties in the obtained dependences are also shown by
dashed lines. 

The profile  of the $H(z)$ function, thus extracted from the analysis of the observational data,
may further be used to compute the temporal behaviour of the FRW scale factor $a(t)$ 
and it's time derivatives
$\dot{a}$, $\ddot{a}$. The steps for this computation are briefly outlined below. 
The FRW scale  $a$ is normalised to unity
at the present epoch ($z=0$) and is 
related to redshift by the relation $1/a = 1+z$. Since $H = \dot{a}/a$, 
this corresponds to $dt = -\frac{dz}{(1+z)H(z)}$ which on integration gives 
\begin{eqnarray}
\frac{t(z)}{t_0} &=& 1 - \frac{1}{ t_0}\int_{z}^0 \frac{dz'}{(1+z')H(z')}\,,
\label{eq:aa1}
\end{eqnarray}
where $t_0$ denotes the present epoch which is also normalised to unity at 
present epoch ($t_0=1$). Using the values of $H(z)$ corresponding
to the $H(z)$ profile as depicted in left panel of Fig.\ \ref{fig:2},
we numerically compute the integral  in the right hand side of Eq.\ (\ref{eq:aa1}). 
Using  Eq.\ (\ref{eq:aa1}) and the equation  $1/a = 1+z$,
we may obtain numerically compute simultaneous
values of $a$ and $t$ at any given redshift $z$ which amounts to
obtaining values of $a(t)$-vs-$t$  eliminating $z$ from  the two equations.
This leads to extraction of the temporal behaviour of the scale factor $a(t)$ from the observational data
for the late time domain of the cosmic evolution accessible through Supernova Ia observations.\\

\begin{figure}[H]
\begin{center}
\centering
\includegraphics[scale=0.49]{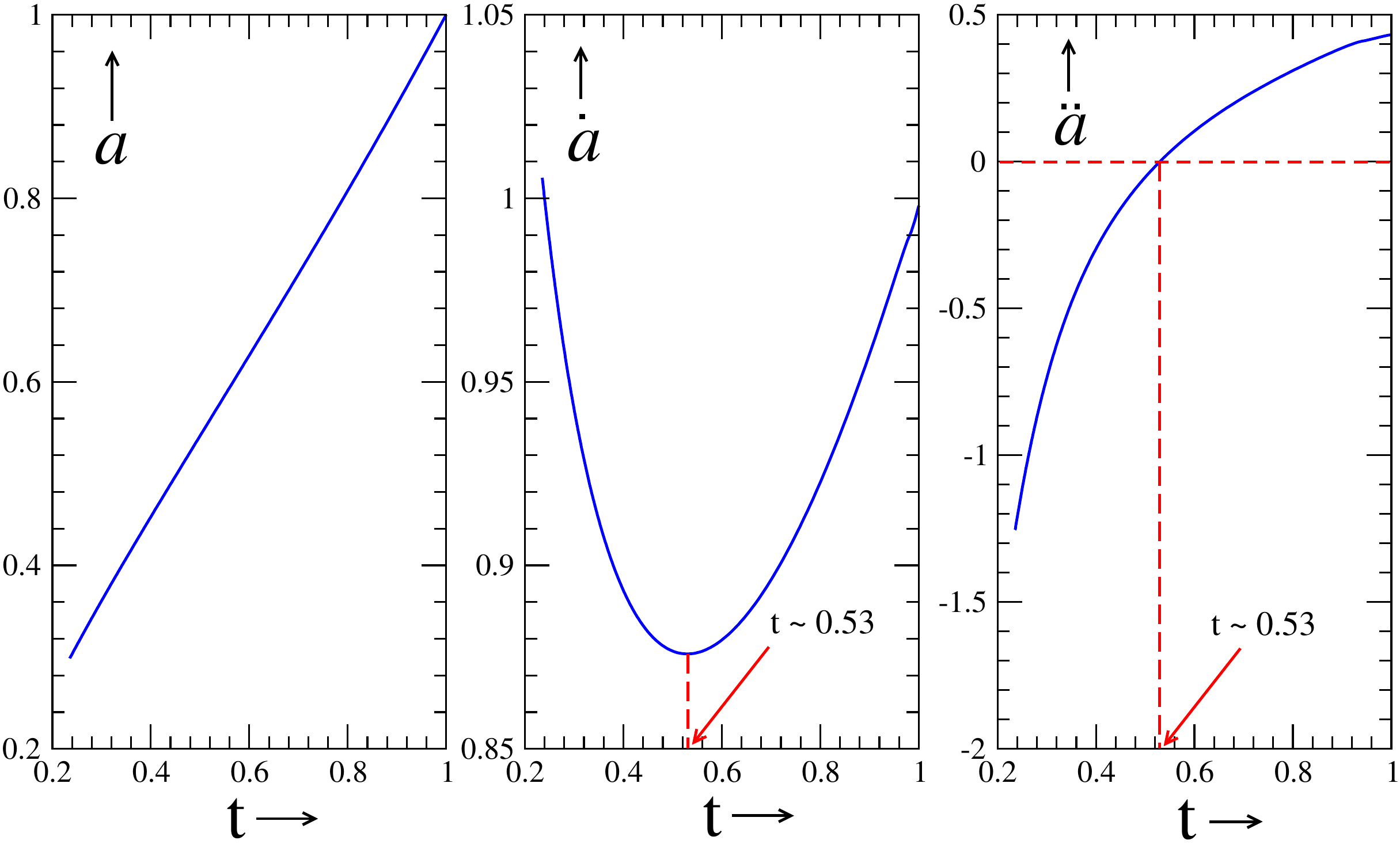}
\caption{Temporal behaviour of scale factor at the best-fit of Pantheon data sets} \label{fig:adot}
\end{center}
\end{figure}

To perform this, we vary $z$  from zero (present
epoch) to $\sim 2.4$ (\textit{i.e.} within the domain of $z$ relevant for Pantheon data set)  
in small steps ($\Delta z = 0.01$). We numerically compute value of the  integral in Eq.\ (\ref{eq:aa1})
 and also the value of $a(z) =
1/(1+z)$  at each $z$-step,  
to obtain the sets of values ($t(z), a(z)$) at  each step.
Using the normalisation of scale factor $a(t)$ as $a=1$
at present epoch ($z=0$ or $t=1$), we found 
that the  range $0<z<2.3$ corresponds 
to $t$-range: $1 > t(z) >0.23$. Values of
 ($t(z), a(z)$) over the domain $0<z<2.3$ provides
 temporal behaviour of scale factor over the time range $0.23<t<1$. 
 Using the obtained profile of $a(t)$, we also obtain a profile
 of $\dot{a}(t)$ and $\ddot{a}(t)$ using direct numerical differentiation techniques.
 We have shown the plots of the profiles $a(t)$,  $\dot{a}(t)$ and $\ddot{a}(t)$
 respectively in the left, middle and right panels of Fig.\ \ref{fig:adot}.
The appearance of a minima at $t\sim 0.53$ in the time-profile of $\dot{a}$ (middle panel)
or  equivalently,the corresponding  change of sign of $\ddot{a}$ in the plots of right panel figure
clearly signifies the
transition from decelerated to an accelerated phase of expansion
during the late time of cosmic evolution.

\section{Results and Discussion}
\label{Sec:IV} 
In this section we present the results regarding the observational constraints on
non-minimally coupled curvature-matter models for two types of
fluid pressure scenarios - the exponential and power law profiles as discussed in
Sec.\ \ref{Sec:II}. The energy density of the fluid
in the context of the two scenarios are given by 
Eqs.\ (\ref{eq:rhoexp}) and Eq.\ (\ref{eq:b6})
respectively.  The temporal evolution of the energy 
density $\rho(t;\lambda, n,k)$ 
has been expressed in terms of
 cosmological quantities like $a,\dot{a},H,R$ 
\textit{etc}. The expression also involves  
the dimensionless parameters: $\lambda, n, k$. Temporal profile of 
the cosmological quantities during the late time phase of the cosmic 
evolution have been extracted from observational data and 
presented in Sec.\ \ref{Sec:III}.
The constant $\lambda$ gives the strength of the non-minimal
coupling between matter and curvature sector. The parameter  $k$  is   
involved in the
modelling of fluid pressure and $n$ denotes the power 
of the gravity term ($f_2(R) = R^n$)
that is coupled to the matter lagrangian. The energy density of a fluid
always being a positive quantity, the constraints  
on parameters $(\lambda, n,k)$ 
for such models come from imposition of the constraint 
$\rho(t; \lambda, n,k) >0$ for all $t$, where $\rho$ at any $t$ is evaluated 
with the values of cosmological quantities ($a,\dot{a},H,R$) 
at that $t$ as obtained
from the analysis of observational data. \\

We presented the observational constraints on the parameters $(\lambda,n,k)$ 
by depicting the allowed area of the $k$-$n$ parameter space for certain chosen
values of the coupling parameter $\lambda$. Obtaining these allowed regions required a thorough scanning of the parameter space and
the pragmatic approach involves predefining a range for the parameters $n$ and $k$ that requires scanning, alongside selecting a scope for the parameter $\lambda$.
In this study, we purposefully choose the relevant range   of 
the coupling strength $|\lambda|$ spanning from 0.1 to 10. 
This choice is motivated by  our intention to comprehensively investigate the cumulative impact of all terms  involved in the primary action. 
When the coupling strength takes on higher values ($|\lambda| > 10$), the significance of $f_1(R) = R$ diminishes compared to the prevalence of other terms. Conversely, for lower coupling values ($|\lambda| < 0.1$), the impact of non-minimal coupling becomes overshadowed by the prominence of other terms within the action. 
We, thus, purposefully choose to focus on benchmark values of $\lambda$ that fall within the above mentioned range in order to illustrate our findings.
When determining the range of parameter space  to be scanned
 for obtaining observationally allowed ranges of $n$ and $k$,
 we systematically investigated various distinct regions within the parameter space. 
In the parameter space domain  corresponding to  
 exceedingly high or low values of $n$ and $k$ (beyond the limits of $-10 \leqslant n \leqslant 10$ and $-20 \leqslant k \leqslant 20$), we did not observe any distinctive, disconnected and closed domains which are observationally allowed 
 or disallowed, that could provide novel and significant insights into those specific portions of the parameter space.  
 Hence, in   presenting the outcomes   in Fig. 4, we have depicted a confined region within the parameter space, defined by  $-10 \leqslant n \leqslant 10$ and $-20 \leqslant k \leqslant 20$. Within this defined realm, conspicuous patterns of various disallowed domains have come to light, providing a scope to comprehend the interrelations between the parameters $n$ and $k$ in congruence with  observations.\\

\begin{figure}[H]
\centering
\begin{subfigure}[b]{0.49\textwidth}
\includegraphics[width=\textwidth]{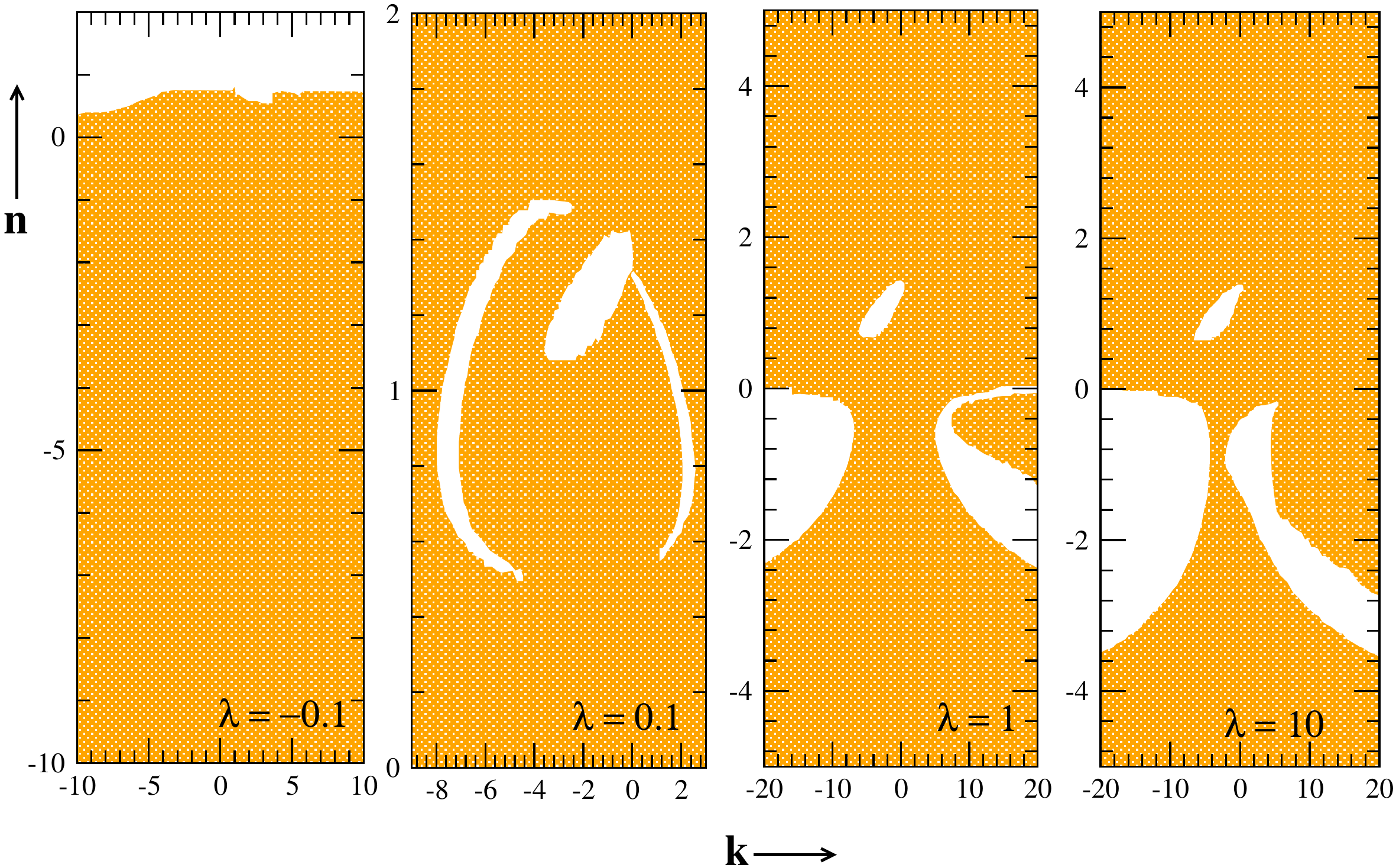}
\end{subfigure}
\begin{subfigure}[b]{0.49\textwidth}
\includegraphics[width=\textwidth]{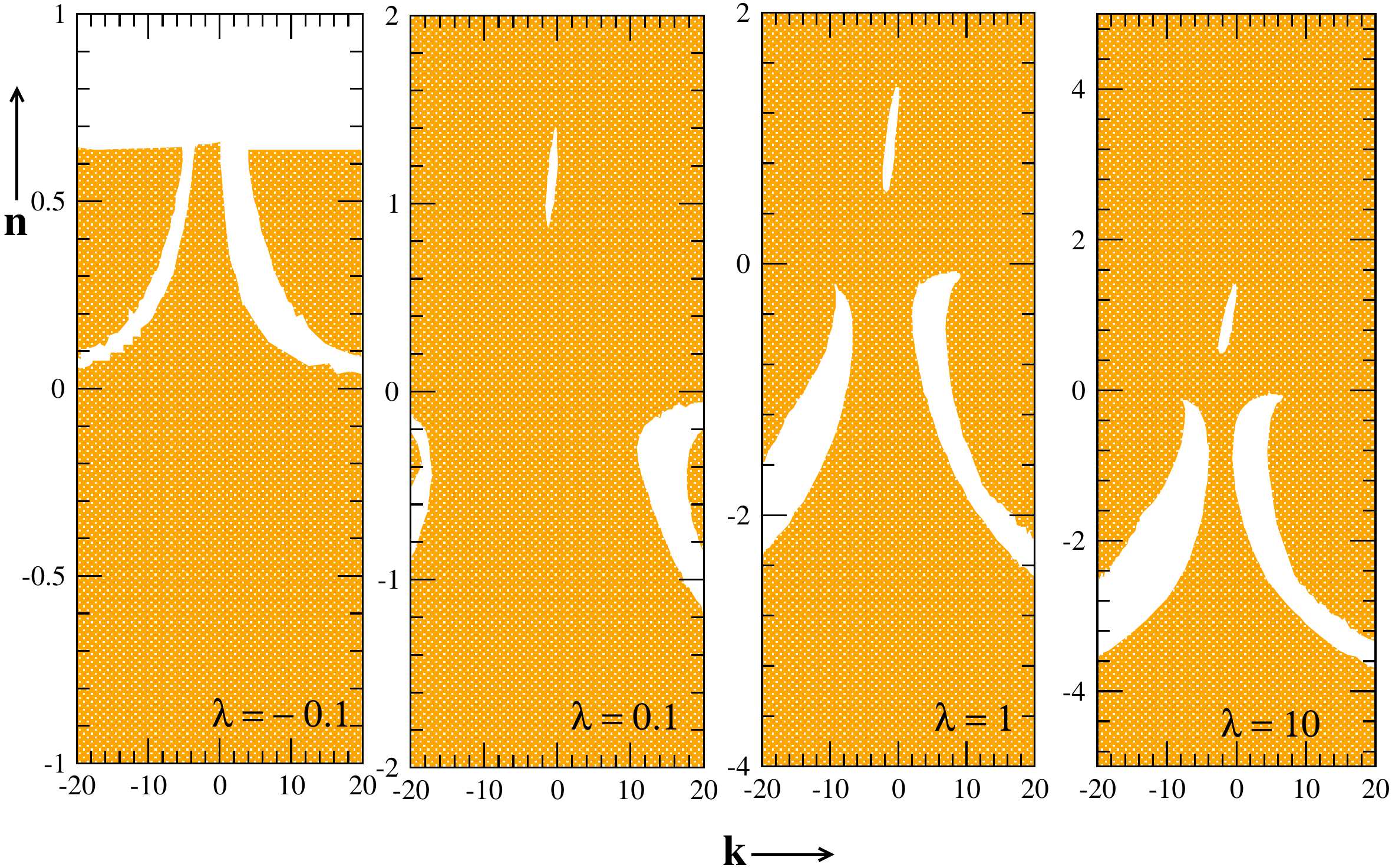}
\end{subfigure}
\caption{Left Panel: Allowed region of $k$-$n$ parameter space for $\lambda = -0.1, 0.1, 1, 10$ for \textit{exponential}
  model (discussed in text).  Right Panel: Allowed region of $k$-$n$ parameter space for $\lambda = -0.1, 0.1, 1, 10$ for \textit{power law}
  model (discussed in text)} 
\label{fig:p3}
\end{figure}

We presented the
results for both the fluid pressure models (exponential and power law) for four
benchmark values: $\lambda = -0.1, 0.1, 1, 10$. 
The allowed
regions in the parameter space,
 which correspond to $\rho >0$, are shown
by shaded regions  in Fig.\ \ref{fig:p3}. The left and right panels of the figure 
correspond to the parameter space constraints obtained for
exponential and power-law fluid-pressure models respectively.
We observe from Fig.\ \ref{fig:p3} that, for both classes of fluid pressure
models, a wide region of the explored portion of the parameter space is
allowed which in turn implies that the Supernova Ia data (Pantheon) and Observed Hubble data
robustly allows non-minimally coupled matter-curvature
scenarios. However for negative values of $\lambda$,
 the positive values of the $n$ (which is the power
of $R$ in the   matter-curvature coupling term) gets severely constrained.\\

\begin{figure}[H]
\centering
\begin{subfigure}[b]{0.49\textwidth}
\includegraphics[height=7cm, width=\textwidth]{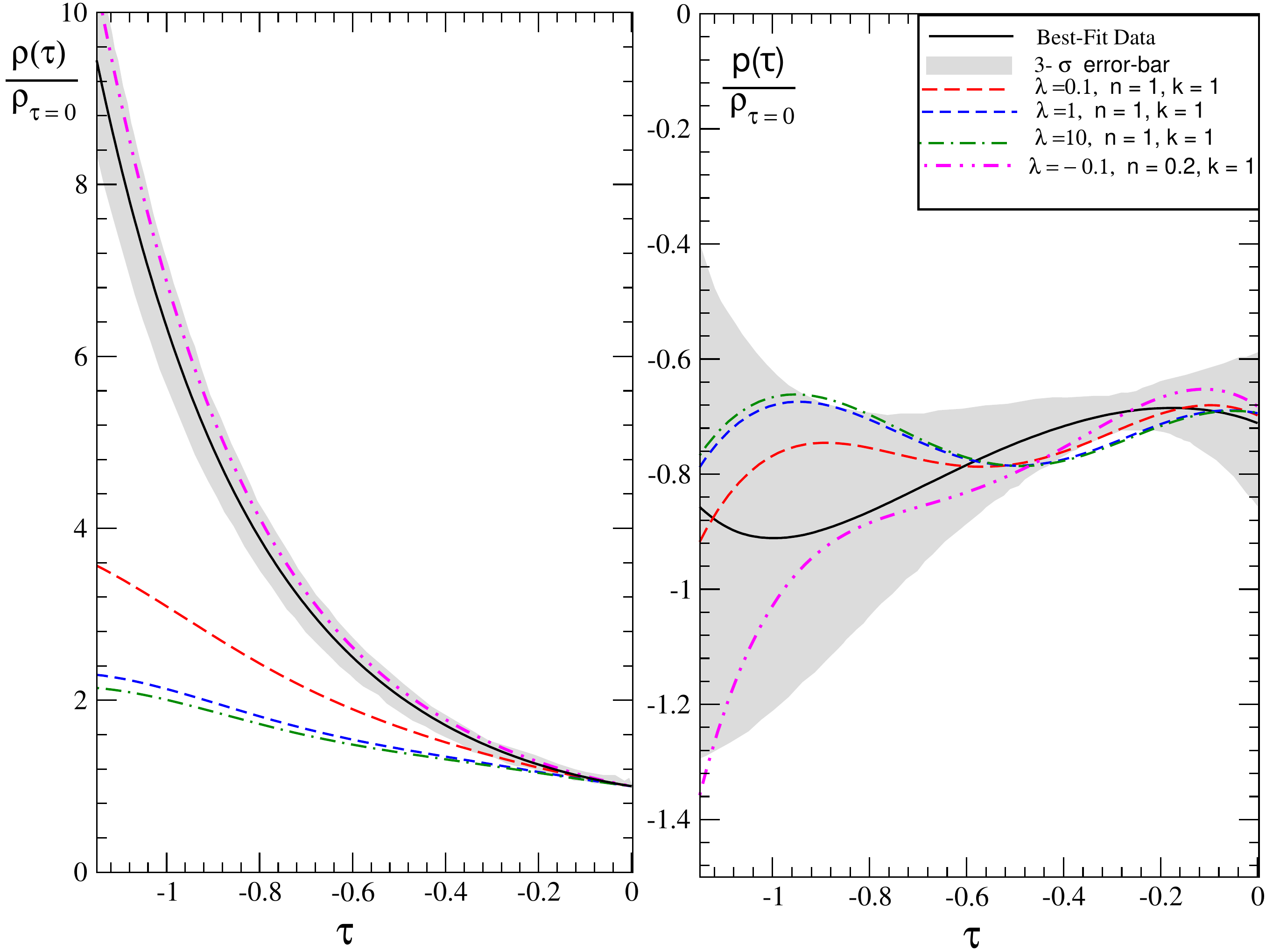}
\end{subfigure}
\begin{subfigure}[b]{0.49\textwidth}
\includegraphics[height=7cm, width=\textwidth]{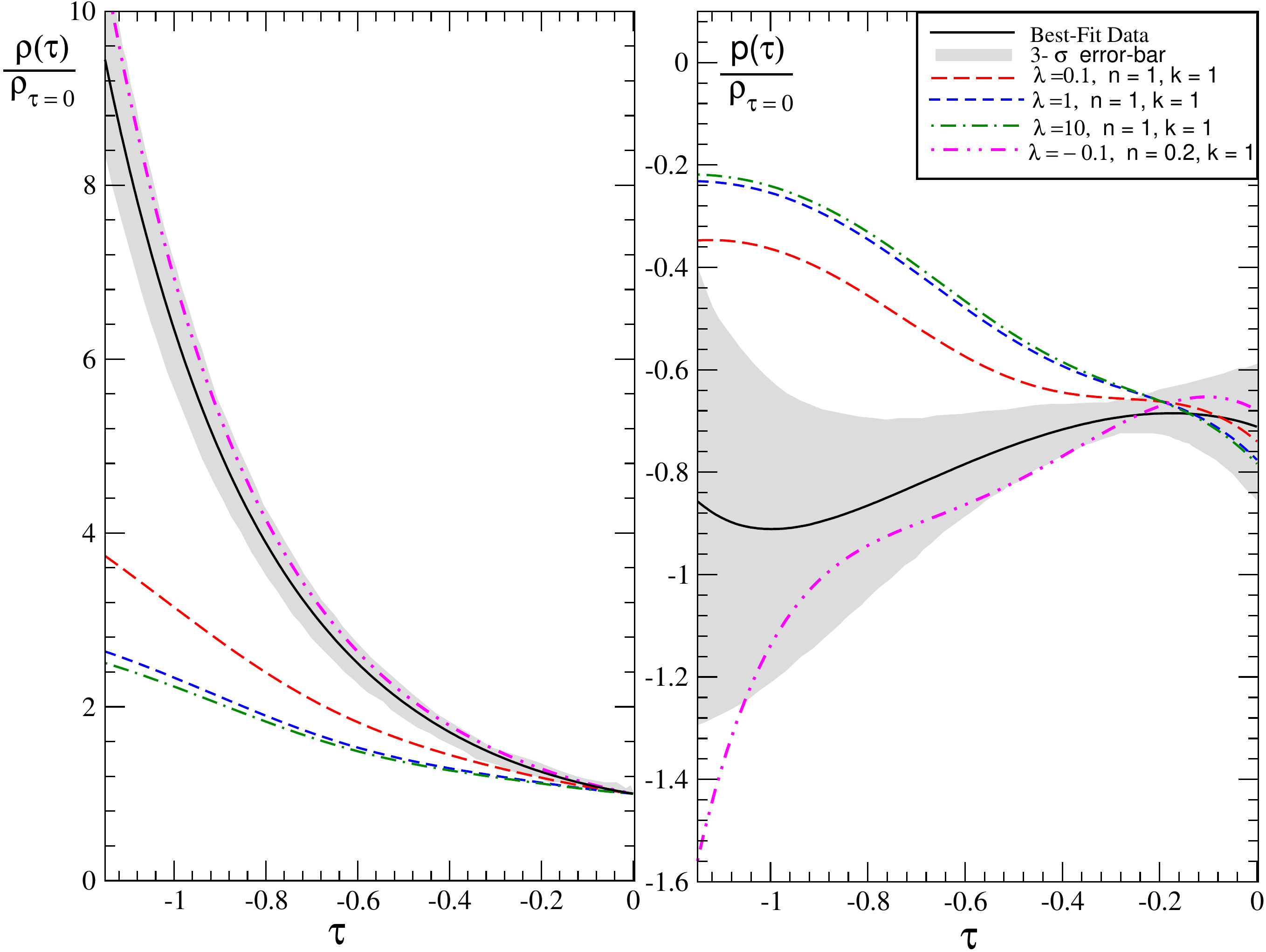}
\end{subfigure}
\caption{Left Panel: Temporal behaviour of energy density (in the left) and pressure (in the right) 
for different 
benchmark values of the parameters $(\lambda,n,k)$ for \textit{exponential}
  model (discussed in text).  Right Panel: Temporal behaviour of energy density (in the left)  
  and pressure (in the right)  for different 
benchmark values of the parameters $(\lambda,n,k)$ for \textit{power law}
 model (discussed in text)  } 
\label{fig:p1a}
\end{figure}

In Eqs.\ (\ref{eq:pexp}) and  (\ref{eq:b5}) 
we have also expressed the temporal profile
of the fluid pressure $p(t; \lambda,n,k)$ 
for exponential and power law models respectively. 
To see the time evolution of energy density 
and pressure in the context of
non-minimally coupled matter-curvature scenarios 
we have shown the plots  of the temporal profile
of energy density and pressure for both fluid  models in Fig.\ \ref{fig:p1a} for
certain benchmark values of the parameter set 
$(\lambda,n,k)$ chosen from the allowed
domains corresponding to both fluid models.

For convenience, we use a dimensionless time parameter $\tau$ 
defined as $\tau = \ln a$ 
to show the time profile of the energy density and pressure in Fig.\ \ref{fig:p1a}.
The late-time domain of cosmic evolution is accessible in 
SNe Ia observations of Pantheon
sample which corresponds to the range $-1.18 \leq \tau \leq 0$ with $\tau = 0$ corresponding
to the present epoch $(a=1)$. We also normalise 
the energy density $\rho(\tau)$ and pressure
$p(\tau)$ by the energy density value at present epoch $\rho_0$ to get rid of dimensions
and plotted $\rho(\tau)/\rho_0$ and $p(\tau)/\rho_0$  as a function of $\tau$ for
different choices of the parameters $k,n,\lambda$.
In the left (right) panel of Fig.\ \ref{fig:p1a} we showed
these plots for  $\rho(\tau)/\rho_0$ and $p(\tau)/\rho_0$ and for the 
exponential (power law) model. All the curves are shown in 
Fig.\ \ref{fig:p1a} correspond to $k=1$.
The 4-curves in each of plots correspond to 4 different choices of the set
$(\lambda,n)$ \textit{viz.} (0.1, 1), (1,1), (10,1), (-0.1,0.2), all of 
which are well within the observationally allowed region of parameter values for 
each of the models as depicted in Fig.\ \ref{fig:p3}. 
For comparison, in these plots, we also showed the best-fit (solid curve) and $3\sigma$ uncertainties
(shaded region) of
the energy density and pressure profiles that result 
from usual analysis of observational data corresponding 
to the $\Lambda-$CDM model.   \\

We see from Fig.\ \ref{fig:p1a} that, for  exponential and power-models
the energy density and pressure profile corresponding
to parameter values ($\lambda = -0.1, n=0.2, k=1$) 
remain mostly well within the $3\sigma$ region
of the corresponding quantities as extracted from the observation.
For other 3 sets of parameter values $(\lambda,n,k)$ considered here, \textit{viz.}
 (0.1, 1,1), (1,1,1), (10,1,1), the energy density curves
 are outside the corresponding 3$\sigma$ range obtained from observation.
A close look
at the plots of pressure profile for exponential and logarithmic models
in Fig.\ \ref{fig:p1a} reveals that the fluid pressure profile
for exponential models also
lies almost entirely within its the 3$\sigma$ observed limits, for 
the full range of time ($\tau$) accessible in SNe Ia observations,
while for power law model,  
the corresponding curve is outside
the 3$\sigma$ observed limits  for a small temporal regime during
a relatively earlier part of late-time cosmic evolution. 
 Therefore, non-minimally
coupled matter-curvature scenarios with fluid pressure modelled as
$p ~\sim e^{ak}$ 
mimics the outcome
of the $\Lambda-$CDM model in terms of energy density and pressure profiles
 within their 3$\sigma$ limits,
for the model parameters taking values in the close proximity
of the values $\lambda = -0.1, n = 0.2, k = 1$. We may mention here,
that the $\Lambda-$CDM model which, though fits the data well, is plagued with
the coincidence and fine-tuning problems, whereas the models with 
non-minimally
coupled matter-curvature scenarios are free from such problems.\\

In the light of the above results we also
investigated the issue  of  energy transfer between curvature and fluid sectors
owing to the considered non-minimal coupling between them.
For the non-minimally coupled curvature-fluid models considered in this article,
 the energy-balance equation takes the form  \cite{Harko:2015pma},
\begin{eqnarray}
\nabla^{\mu}T_{\mu \nu}   & = & \frac{\lambda F_2(R)}{1 + \lambda f_2(R)}\Big{(}g_{\mu \nu} \mathcal{L}_m - T_{\mu \nu}\Big{)} \nabla^{\mu} R   \,.
\label{eq:Q}
\end{eqnarray}
For an ideal fluid with energy density $\rho$ and pressure $p$ in FRW spacetime background
and with $\mathcal{L}_m = p$ as considered in the work,  
the $\nu=0$ component of Eq.\ (\ref{eq:Q}) takes the form
\begin{eqnarray}
\dot{\rho} + 3H(\rho + p) &=&   - \frac{\lambda F_2(R)}{1 + \lambda f_2(R)}\Big{(}  p + \rho\Big{)}   \dot{R}\, 
\equiv  Q  \,.
\label{eq:Q1}
\end{eqnarray}
 Note that,
in absence of any non-minimal coupling $(\lambda = 0)$, the above 
energy balance equation reduces to usual continuity equation $\dot{\rho} + 3H(\rho + p) = 0$
of   FRW universe.
The term $Q$ on right hand side of Eq.\ (\ref{eq:Q1}) is a measure of
rate of energy exchange between matter and curvature sectors.
In the context of Eq.\ (\ref{eq:Q1})  we may mention that,
thermodynamical implications of curvature-matter coupling scenarios at cosmological scales have been investigated
in \cite{Harko:2015pma,Harko:2014pqa,Moraes:2016mlp}, where from the perspective of
thermodynamics of open systems,
the  energy balance equation (\ref{eq:Q1}) has been interpreted  
to describe  particle creation in FRW universe. Such an interpretation
requires   referring  to the term $-Q/3H$ as the creation pressure $p_c$
associated with the particle creation,  so as to cast 
the energy balance equation in the form  $\dot{\rho} + 3 H (\rho + p + p_c) = 0$, with 
creation pressure subject being to the constraint that $p_c \equiv -Q/3H <0$, implying
the requirement that $Q$ always remains positive. \\

\begin{figure}[H]
\begin{center}
\centering
\includegraphics[scale=0.49]{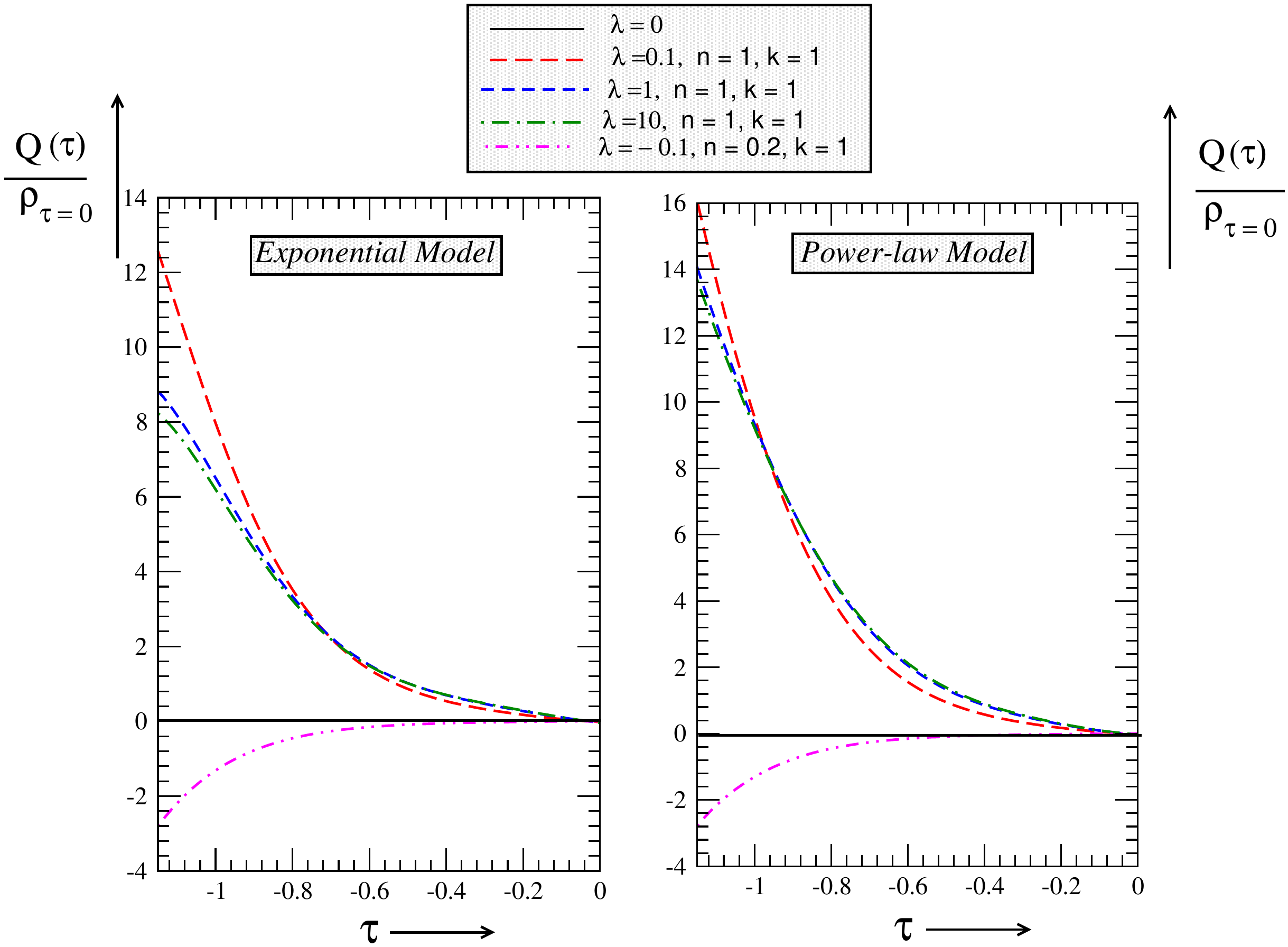}
\caption{Temporal behaviour the function $Q(t;\lambda,n,k)$ (discussed in text)
for different sets of  benchmark values of parameters $Q(t;\lambda,n,k)$
for exponential (left panel) and power-law (right panel) models} \label{fig:p4} 
\end{center}
\end{figure}

In the context of non-minimally coupled curvature-fluid models considered here with
  $f_2(R) = R^n$,
  we may write down the expression for the time profile of the $Q$-function from Eq.\ (\ref{eq:Q1}) 
corresponding to the parameter set $(\lambda,n,k)$ as 
\begin{eqnarray}
Q(t;\lambda,n,k) &=& -\frac{\lambda n R^{n-1}}{1 + \lambda R^n}\Big{[}  p(t;\lambda,n,k) + \rho(t;\lambda,n,k)\Big{]}   \dot{R}
\label{eq:Q2}
\end{eqnarray}
where the energy density   
$\rho(t;\lambda,n,k)$ and fluid pressure 
$p(t;\lambda,n,k)$ profiles corresponding to both exponential and power law models
are given by Eqs.\  (\ref{eq:rhoexp}) - (\ref{eq:b5}).
We use Eq.\ (\ref{eq:Q2}) to compute the function $Q(t;\lambda,n,k)$ for given
choices of $\lambda,n,k$.
Time dependences
of cosmological parameters \textit{viz.} $a$, $H$, $R$ \textit{etc.}, as extracted from observed data
are  instrumental
in the determination of  the  temporal profile of $Q$.
In Fig.\ \ref{fig:p4}
we have shown the time dependence of   $Q$ for both exponential (left panel)
and power law (right panel) models by plotting $Q(\tau)/\rho_0$ vs $\tau$ for the
same 4 benchmark values of parameters ($\lambda,n,k$) as taken earlier.
We observe from Fig.\ \ref{fig:p4} that, 
corresponding to the benchmark cases with $n=1$ (\textit{i.e.} $f_2(R)=R$) and 
$k=1$ (which correspond to $p \sim e^a$ for exponential model 
and $p \sim a$ for power law model), 
$Q(\tau)$ remains always positive.
So, certain domains of   $(\lambda,n,k)$ parameter space 
of non-minimally coupled matter-curvature models are allowed
from the combined analysis of Pantheon data and OHD,
for which the possibility of 
interpreting the energy balance equation in terms of particle creation \cite{Harko:2015pma} 
remains open.
On the other hand, the estimation of $Q(\tau)$ for parameter values
($\lambda = -0.1,n=0.2,k=1$), (which  mimics
energy density and pressure profile obtained from the data analysis with $\Lambda$-CDM
model  as explained earlier) gives negative values of $Q$ for the entire time
range probed in SNe Ia observations. 
We also observe from Fig.\ \ref{fig:p4} that,
the absolute value of the rate of
energy transfer between curvature and matter sectors 
and hence the effect of the considered non-minimal coupling,
monotonically decreases as time approaches towards the present epoch.

\section{Conclusion}
\label{Sec:V} 
In this work we considered 
non-minimally coupled curvature-matter models
of gravity and investigated its cosmological implications 
in the light of  luminosity distance and redshift measurements
of Supernova Ia events.
The non-minimal curvature-matter coupling has been introduced
by adding  
a term $\int  \left[\lambda R^n \mathcal{L}_m\right] \sqrt{-g}d^4x$ to the
usual action for Einstein's gravity involving the
Einstein Hilbert action and minimally coupled matter action.
The parameters $\lambda$ and   $n$   fix the strength and
nature of the non-minimal coupling.
To explore consequences of such non-minimal couplings  
in relation to evolutionary aspects of the universe at large scales,
we considered homogeneous and isotropic spacetime geometry
of the expanding universe described by a
flat FRW metric involving the time dependent scale factor $a(t)$. 
The matter content of
the universe is modelled as a perfect fluid 
characterised by energy density $\rho(t)$
and pressure $p(t)$ and chosen
 the form of matter Lagrangian as $\mathcal{L}_m = p$. Such a choice for  Lagrangian density correctly reproduces the hydrodynamical equations for the perfect fluid and also may cause to vanish the extra force owing to departure from motion along the geodesic arising from a non-vanishing covariant derivative of energy-momentum tensor in coupled scenarios. For such considerations, the field equations corresponding to the modified action containing non-minimal term take forms exhibiting connection between
the $a(t),\rho(t),p(t)$ through 
their time derivatives and various other functions 
like the Hubble function $H(t)$ and the Ricci scalar $R(t)$.
The functions $H(t)$, $R(t)$ are however also directly related
to the scale factor $a(t)$ and its derivatives.
The evolution equations also involve the parameters $\lambda$ and $n$
related to the curvature-matter coupling and the minimal coupling scenario
corresponds to $\lambda = 0$, for which the equations reduce
to the   Friedman equations.  
Unlike Friedman's equations, 
their corresponding analogues which govern 
the evolutionary dynamics 
in presence of non-minimal curvature-matter coupling,
involves time derivates of fluid pressure $p$. 
We investigate the observational constraints on 
the non-minimal models choosing two different ansatzes for
$p(t)$ referred to in the paper as the `exponential model'
and `power law model', where the temporal profile of the
fluid pressure has been
parametrized in terms of a dimensionless parameter $k$
as $p \sim e^{ka}$ and $p \sim a^k$ respectively. 
Consequently, in the context of this work,
the interplay of the three parameters
$(\lambda,n,k)$ plays a pivotal role in testing the consistency
of non-minimally coupled fluid-curvature scenarios
with the observed data.\\

From a comprehensive analysis of 
 Pantheon compilation of 1048 SNe Ia data points 
and 54 data points on measurement of Hubble parameter from OHD,
we obtained time dependences of the relevant cosmological
parameters $a(t),H(t),R(t)$ and  their time derivatives over
the late time phase of cosmic evolution.
Taking into account all these information and using the evolution
equations for non-minimally coupled scenarios,
we numerically computed the energy density $\rho(t,\lambda,n,k)$ and pressure $p(t,\lambda,n,k)$
profiles for different values of $(\lambda,n,k)$ thoroughly scanning
a portion of the parameter space: [$-0.1 \leqslant \lambda \leqslant 10 ; -10 \leqslant n \leqslant 10;
-20 \leqslant k \leqslant 10$]. 
We also obtained and presented in Fig.\ \ref{fig:p3}, 
the regions in the parameter space corresponding to
parameter values  giving $\rho(t,\lambda,n,k) > 0$ for all $t$ which is a 
trivial but essential
requirement for viability of cosmological models.
We found there exist  large domains in the
$(\lambda,n,k)$-parameter space for which models with non-minimal 
curvature-matter  mixing
stand  as  viable cosmological models  reproducing 
the observed features of late time  cosmic evolution.
We have also seen that there exist a small range of parameter values around
($\lambda = -0.1,n=0.2,k=1$) for which the computed temporal profiles of
$\rho$ and $p$ mimic the  corresponding profiles obtained from
the analysis of the data using $\Lambda-$CDM model 
in the context of usual minimal coupling scenario. \\

The energy-momentum tensor has a non-vanishing covariant derivative 
($\nabla^\mu T_{\mu\nu} \neq 0$) in the realm of non-minimally coupled curvature-matter
scenarios and this implies 
exchange of energy between curvature and matter sectors. 
We found that, the absolute
value of the rate of energy transfer between the two sectors  
monotonically decreases as time approaches towards the
present epoch.
This  dynamics of energy exchange may be expressed 
by an equation portraying the energy balance between the two sectors
in FRW background 
as $\dot{\rho} + 3H(\rho + p + p_c) = 0$, with the function $p_c(t)$ (multiplied by $3H$) 
providing a measure of the  rate of energy transfer owing to the non-minimal
coupling between curvature and matter.
From the perspective of thermodynamics of open systems, as extensively discussed  in 
 \cite{Harko:2015pma,Harko:2014pqa,Moraes:2016mlp},
curvature-matter coupling allows   production of 
a substantial amount of comoving
entropy during late time evolutionary phase of the universe. This opens up
the possibility to interpret the energy-balance equation describing
particle creation in FRW universe, with  $p_c$ realised as  
the (creation) pressure related to the particle creation. Such an interpretation
however requires $p_c$ to be negative. Our investigation shows,
there exist values of model parameters $(\lambda,n,k)$, 
for which the requirement $p_c<0$ (for all time) is met.
Thus the `particle creation' interpretation of energy balance equation
in non-minimally coupled curvature-matter scenarios
remains an open possibility in the context of SNe Ia data.

\paragraph{Acknowledgement}\
We express our gratitude to the referees for providing valuable suggestions. A.C. would like to thank Indian Institute of Technology, Kanpur for supporting this work by means of Institute Post-Doctoral Fellowship \textbf{(Ref.No.DF/PDF197/2020-IITK/970)}. 
 

\end{document}